\documentclass[12pt]{article}

\usepackage{times}
\usepackage{geometry}
\usepackage{graphicx}
\usepackage[utf8]{inputenc}
\usepackage{enumitem,amssymb}
\usepackage{ragged2e}
\newlist{thematic}{itemize}{8}
\setlist[thematic]{label=$\square$}
\usepackage{pifont}
\newcommand{\cmark}{\ding{51}}%
\newcommand{\done}{\rlap{$\square$}{\raisebox{2pt}{\large\hspace{1pt}\cmark}}%
\hspace{-2.5pt}}

\usepackage{hyperref,xcolor,amsmath}
\usepackage[superscript,biblabel]{cite}
\usepackage[super,sort&compress]{natbib}
\usepackage{aas_macros}
\usepackage[font={it},labelfont=bf]{caption}
\usepackage{wrapfig,sidecap}

\begin{document}
\begin{raggedright}
\huge
Astro2020 Science White Paper \linebreak

Measuring the Hubble Constant Near and Far in the Era of ELT's\linebreak
\normalsize

\noindent \textbf{Thematic Areas:} \hspace*{60pt} $\square$ Planetary Systems \hspace*{10pt} $\square$ Star and Planet Formation \hspace*{20pt} \linebreak
  $\square$ Formation and Evolution of Compact Objects \hspace*{31pt} 
  $\done$ Cosmology and Fundamental Physics \linebreak
  $\square$  Stars and Stellar Evolution \hspace*{1pt} 
  $\done$ Resolved Stellar Populations and their Environments \hspace*{40pt} \linebreak
  $\square$  Galaxy Evolution   \hspace*{45pt} 
  $\done$ Multi-Messenger Astronomy and Astrophysics \hspace*{65pt} \linebreak
  
\textbf{Principal Author:}

Name: Rachael L. Beaton 
 \linebreak						
Institution: Princeton University \& Carnegie Observatories
 \linebreak
Email: rbeaton@princeton.edu
 \linebreak
Phone: 434 760 1404 (cell)
 \linebreak
 
\textbf{Co-authors:} \linebreak
Simon Birrer, UCLA (Primary Co-Author; sibirrer@astro.ucla.edu); \linebreak
Ian Dell'Antonio, Brown (ian\_dell'antonio@brown.edu); 
Chris Fassnacht, UC Davis (cdfassnacht@ucdavis.edu); 
Danny Goldstein, CalTech (danny@caltech.edu);
Chien-Hsiu Lee, NOAO (lee@noao.edu); 
Peter Nugent, LBNL (penugent@lbl.gov);
Michael Pierce, U Wyoming (mpierce@uwyo.edu);
Anowar J. Shajib, UCLA (ajshajib@astro.ucla.edu);
Tommaso Treu, UCLA (tt@astro.ucla.edu). \linebreak

\end{raggedright}

\smallskip
\noindent \textbf{Abstract:} \linebreak
% RLB -- new catchier abstract aimed less specifically for ELT's
Many of the fundamental physical constants in Physics, as a discipline, are measured to exquisite levels of precision. 
The fundamental constants that define Cosmology, however, are largely determined via a handful of independent techniques that are applied to even fewer datasets.
The history of the measurement of the Hubble Constant (H$_0$), which serves to anchor the expansion history of the Universe to its current value, is an exemplar of the difficulties of cosmological measurement; indeed, as we approach the centennial of its first measurement, the quest for H$_0$ still consumes a great number of resources. 
In this white paper, we demonstrate how the approaching era of Extremely Large Telescopes (ELTs) will transform the astrophysical measure of H$_0$ from the limited and few into a fundamentally new regime where
(i) multiple, independent techniques are employed with modest use of large aperture facilities and 
(ii) 1\% or better precision is readily attainable.
This quantum leap in how we approach H$_0$ is due to the unparalleled sensitivity and spatial resolution of ELT's and the ability to use integral field observations for simultaneous spectroscopy and photometry, which together permit both familiar and new techniques to effectively by-pass the conventional ``ladder'' framework to minimize total uncertainty. 
Three independent techniques are discussed -- (i) {\it standard candles} via a two-step  distance ladder applied to metal, poor stellar populations, (ii) {\it standard clocks} via gravitational lens cosmography, and (iii) {\it standard sirens} via gravitational wave sources -- each of which can reach 1\% with relatively modest investment from 30-m class facilities. 

\pagebreak

\section{Context}
Since its theoretical prediction\cite{lemaitre_1927} and experimental discovery\cite{hubble_1929}, the Hubble constant (H$_0$) has been a critical parameter for cosmological models. 
The history of its measurement is, indeed, demonstrative of its importance, as the resolution of controversy in its measured or inferred value from independent lines of evidence has led to fundamental discoveries in cosmology, including most recently Dark Energy. 
Moreover, the pursuit of ever more robust measurements of H$_0$ has motivated facilities and  refinements of instrumentation and technique that have influence across astrophysics. 

Since the conclusion of the HST Key Project in 2001\cite{freedman_2001}, the landscape for measuring H$_0$ has evolved dramatically, culminating in its measure at 2.3\%\cite{riess_2016,riess_2018a,riess_2018b} via the traditional Cepheid-based distance ladder. 
Likewise, its measurement via modeling of the anisotropies of the Cosmic Microwave Background (CMB) has also improved from the final results of WMAP\cite{bennett_2013} in 2013 to a 0.6\% measure from Planck in 2018\cite{planck_2018} (TT,TE,EE+lowE+lensing+BAO assuming a flat $\Lambda$CDM model).
The past decade has also seen the experimental realization of long proposed techniques\cite{refsdal_1964,schutz_1986} to measure H$_0$, including gravitational lens cosmography\cite{T+M16,birrer_2019,suyu_2017} and gravitational waves\cite{abbott_2017}, both of which are delivering comparable accuracy and precision to the traditional methods.

As we look toward the 2020's, we do so at yet another conflict in the value of H$_0$\cite{freedman_2017}. 
Currently, the two most precise measurements -- that from the classical Cepheid-based distance ladder\cite{riess_2018a,riess_2018b} and that inferred by modelling the anisotropies in the CMB\cite{planck_2018} with the standard $\Lambda$-CDM cosmology, disagree by more than 3.8$\sigma$. 

The tension is more complex than a simple disagreement between methods.
Independent measurements of H$_0$ in the ``local'' Universe produce values in agreement with the distance ladder.\cite{birrer_2019,abbott_2017}
Recent investigations have shown that, if calibrated either locally or to the CMB, the two tracers of evolution of the expansion, the Baryon Acoustic Oscillations (BAO) and Supernovae Ia, produce results that are largely in agreement.\cite{aylor_2018}
Thus, while the middle-ages of the Universe are well probed by current techniques, how they are anchored -- either in the Universe's youth or at its current age, result in different cosmologies due to the strong degeneracy between H$_0$ and other cosmological parameters. 

Theoretical means to resolve the tension require ``new physics."
Proposed modifications to the standard model include evolving dark energy\cite{poulin_2018}, interacting dark matter\cite{divalentino_2017,divalentino_2018}, and interacting neutrinos\cite{kreisch_2019}, among others.
Moreover, despite an ever-increasing volume of work presenting detailed tests, it remains unclear if there are lingering {\it instrumental systematic effects} within the Planck observations -- which will be tested in the coming years by a suite of independent, high resolution CMB experiments \cite{abazajian_2016,ade_2019}, or if there are nefarious {\it astrophysical systematic effects} within the distance scale -- which are currently being tested via independent standard candles.\cite{beaton_2016}

While the aforementioned on-going studies may indeed provide resolution to the current H$_0$ controversy, the community is still, effectively, limited to two high-precision techniques for measuring H$_0$.
Because H$_0$ is a fundamental quantity, it must be measured rigorously by independent techniques, independent teams, and independent datasets. 
{\bf In this white paper, we highlight the key science contributions that Extremely Large Telescopes (ELT's) will provide to enable three independent and fundamentally different measurements of H$_0$ at the 1\% uncertainty level.}
The different means of measuring H$_0$ for ELT's are: (i) using {\it standard candles} via luminosity distances (\S~\ref{sec:candles}), (ii) using {\it standard clocks} via gravitational lens time delays (\S~\ref{sec:clocks}), and (iii) using {\it standard sirens} via gravitational wave sources (\S~\ref{sec:sirens}).

\section{H$_0$ via Standard Candles} \label{sec:candles}

The modern distance ladder,\cite{riess_2009} combines geometric calibrations of Cepheid type variables, subsequent calibration of SNe~Ia, and determination of H$_0$ from SNe~Ia in the Hubble Flow. 
The end uncertainties are a combination of measurement uncertainties and the astrophysical variance of each stage, such that the most effective means of reaching higher precision is to eliminate steps in the ladder.
Set on a robust foundation of geometric calibration in the Local Group, the ELT's will permit the direct, and routine, measurement of distances to galaxies in the Hubble Flow via the infrared tip of the red giant branch (IR-TRGB).

The end of the red giant branch (RGB) sequence occurs when the He-core reaches a sufficient temperature to lift degeneracy such that the star undergoes the rapid evolution caused by the He-flash;\cite{cassisi_2013,serenelli_2017} the result is that the RGB ends abruptly in color-magnitude space.\cite{lee_1993} 
Because the core reaches a set temperature, the bolometric flux from the core is constant and the resulting luminosity measured in a specific bandpass is determined by how the composition of the stellar atmosphere (metallicity) absorbs and redistributes the flux.\cite{cassisi_2013,serenelli_2017} 
A major advantage of the TRGB-method is that RGB stars can be found in low-stellar density, low-reddening, and low-metallicty stellar halos of galaxies, which, taken together, eliminate many of the lingering terms in the error-budget for Cepheids (crowding, internal extinction, metallicty)\cite{freedman_2010} -- terms that will not necessarily be solved by ELT's.
Cepheids can only be found in star-forming galaxies; in contrast, RGB stars are present in galaxies haloes for all Hubble types. 
Lastly, RGB stars are non-variable, meaning that only a single set of imaging in two bands are required to make the measurement.

\begin{SCfigure} %%%%%%%%%%%%%%%%%%%%%%%%%%%%%%%%%%%%%%%%%%%%%%%%%%%%%%%%%%%%%%%%%%%%%%
\centering
\includegraphics[width=0.60\textwidth]{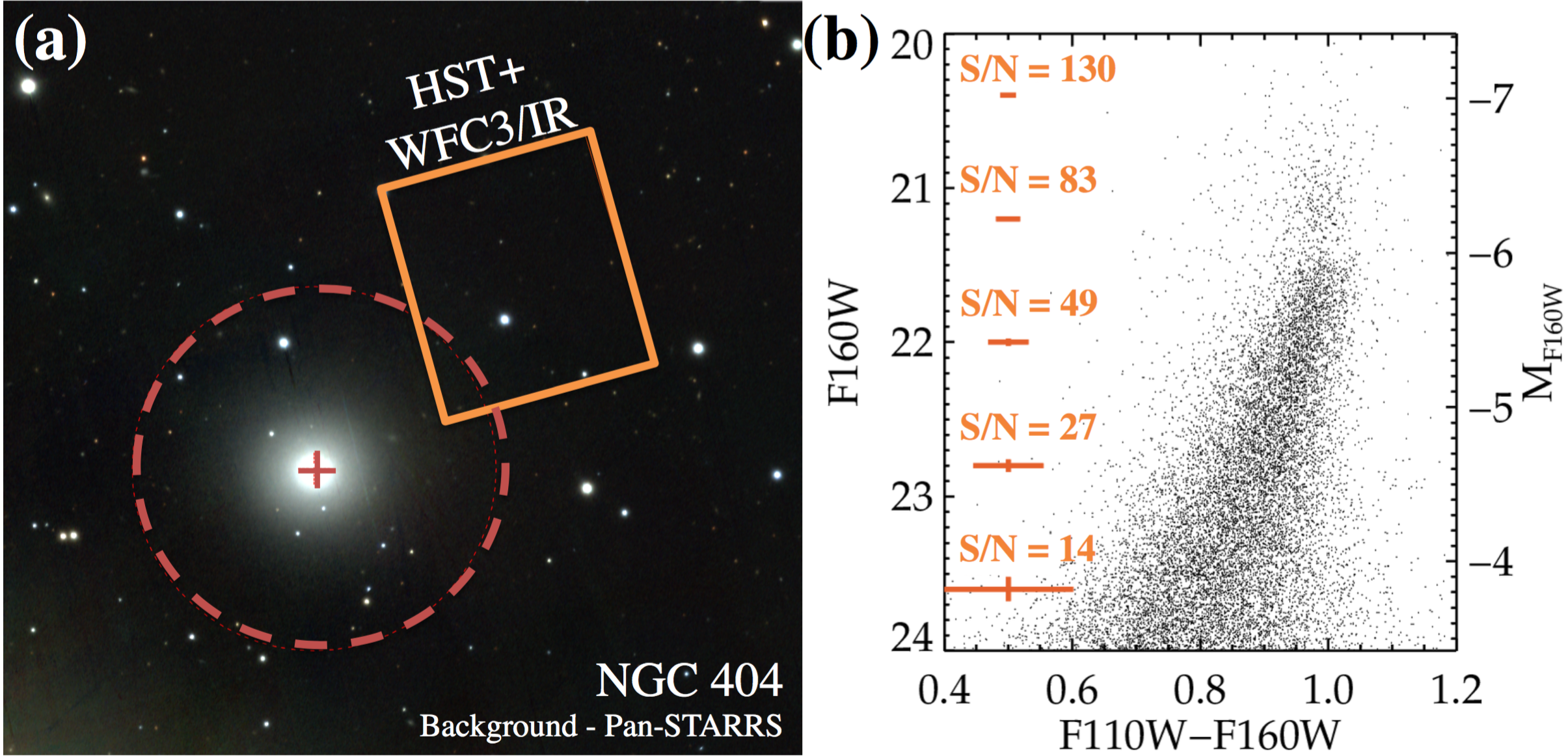}
\caption{ 
Left: Pan-STARRS image of NGC\,404 with the HST+WFC3/IR footprint targeting its ``halo'' overlaid (orange). 
Right: Infrared color-magnitude diagram for NGC\,404\cite{dalcanton_2012} that demonstrates its strong red giant branch. The TRGB is visible at $M_{F160W}\sim-6$~mag and can be measured using edge-detection algorithms.}
\end{SCfigure} %%%%%%%%%%%%%%%%%%%%%%%%%%%%%%%%%%%%%%%%%%%%%%%%%%%%%%%%%%%%%%%%%%%%%%

A full calibration of the SNe~Ia using the optical TRGB is well underway\cite{beaton_2018} and has already produced individual distances at \textless~5\% precision for five galaxies within 20 Mpc\cite{jang_2018,hatt_2018a,hatt_2018b} using a set of rigorous techniques.\cite{hatt_2017, jang_2018} 
Parallel work has established the infrared TRGB (IR-TRGB) as an equally precise distance indicator\cite{madore_2018,hoyt_2018} and at M$_{F160W}\sim-6.0$~mag (similar to $H$), the IR-TRGB is as bright as a 10-day Cepheid (the faintest used for distance determination), but requires only a single epoch of imaging. 
An example of the IR-TRGB is given in Fig.~1, where HST$+$WFC3/IR imaging in the outskirts of the local galaxy NGC\,404 is used to determine its distance by detection of the apparent ``end'' of the RGB sequence.
The most precise distances, \textless5\% precision, require high signal-to-noise photometry to apply a color-magnitude correction,\cite{madore_2018,hoyt_2018,dalcanton_2012} but for 5\% precision, 20$\sigma$ photometry at the tip is sufficient and has been used heavily to map the local cosmic flows.\cite{tully_2016}

The IR-TRGB can measure distances to galaxies directly in the Hubble Flow (e.g., $D \sim$ 100 Mpc) on 30-m class facilities. 
For a galaxy at 100 Mpc ($m-M=35$~mag), the apparent magnitude of the TRGB is m$_H$=29 mag -- as an example, this corresponds to a $\sim$1 hour integration with TMT+IRIS for photometry at 20$\sigma$ ($\sim$0.05 mag uncertainty per star). 
At this distance, each galaxy is an independent, 5\% measurement of H$_0$ and reaching 1\% precision in H$_0$ would naively require $\sim$25 galaxies. 

ELTs can also solve the lingering issues of the \textit{absolute foundation} of the distance scale. 
\emph{Gaia} will provide parallaxes within the Milky Way, but calibration of distance indicators in other local galaxies remains important to fully sample astrophysical variation. 
To date geometric distances exist for only a handful of galaxies, with distances determined from eclipsing binaries being the most widely applicable technique. 
Eclipsing binaries can be identified at 1~Mpc ($\sim$Andromeda distance) in time series imaging on 4 to 8m class facilities,\cite{lee_2014} but the required spectroscopic follow-up to constrain orbital parameters is limited on 10-m class facilities.\cite{vilardell_2010,lee_2014}
With eight binaries, a 2\% distance was determined to the Large Magellanic Cloud\cite{pietrzynski_2013} with additional systems possibly attaining   1\% in the near future.\cite{graczyk_2018}
ELTs will be able to produce distances of this quality with observations of EBs in galaxies across the Local Group to establish a broad and stable foundation for the distance ladder.

%%%%%%%%%%%%%%%%%%%%%%%%%%%%%%%%%%%%%%%%%%%%%%%%%%%%%%%%%%%%%%%%%%%%%%%%%%%%%%%%%%%%%%%%%%%%%%%%%%%%%%%%%%%%
\section{H$_0$ via Standard Clocks} \label{sec:clocks}

An alternative method to determine H$_0$ is gravitational lensing time-delay cosmography.
If the lensed object in a multiply-image strong lensing system has intrinsic variability, then the same variable behavior will appear in the individual lensed images at delayed times due to different light travel paths. 
The time-delay, or difference in travel time, depends on the space-time curvature and the distances involved in the lensing system with the result that with the time delays measured, we can infer absolute distance ratios and measure H$_0$. 
Refsdal\cite{refsdal_1964} first suggested this method using lensed SNe, but this has proven unfeasible due to the rarity of such events. 
Instead, the H0LiCOW\cite{suyu_2017} team has successfully used lensed quasars; using three quadruply lensed quasars and one doubly lensed quasar have determined H$_0$ to 3\% (examples given in Fig.~2a).\cite{birrer_2019}

To use gravitational lensing time-delay for H$_0$ measurements, there are three important ingredients:
(1) \textit{accurate time-delay measurements} -- requiring high cadence, high signal-to-noise photometric monitoring over several months with medium class telescopes for quasar lenses;\cite{courbin_2018}
(2) \textit{a precise mass model for the primary lensing galaxy} -- requiring imaging and spatially resolved 2-D kinematic maps at high angular resolution over the extent of the lens-system, 
and 
(3) \textit{an accounting of the mass distribution along the line-of-sight} -- requiring redshifts and potentially kinematic properties for any nearby galaxies that cause perturbations on the mass model down to the percent level. 
The 2-D kinematic maps of the lensing galaxy are particularly important in the mass modeling to (1) break the mass-anisotropy degeneracy, (2) place valuable constraints on the inner mass profile of the lensing galaxies and thus mitigate the impact of the mass-sheet degeneracy\cite{falco_1985} and (3) reduce the uncertainties on the angular diameter distance to the lens from $\sim20\%$ to $\sim7-8\%$.\cite{shajib_2018}

While current H$_0$ measurements use quasars, the advent of large-sky time-domain surveys (e.g. Pan-STARRS, ZTF, and LSST) makes it now possible to detect lensed supernovae, in particular those of type Ia (SNe~Ia). 
The standard candle nature of SNe~Ia provides a great leverage to gravitational lensing time-delay analysis for cosmography because its well-constrained luminosity provides a natural means to break mass-sheet degeneracy that lensed quasar systems often suffer.\cite{falco_1985}
To use a lensed supernova, high S/N light curves are required for each of the lensed supernova images. 
The image separation may be small, $\sim$0.5$''$ for iPTF16geu\cite{goobar_2017} (Fig.~2a), hence time-series imaging with adaptive optics is essential to obtain the time delays -- only the ELTs can provide sufficient spatial resolution.

The lensing systems will be discovered by current and future deep imaging surveys on 8-meter class telescopes, such as the on-going Hyper Supreme Cam Strategic Survey Program\cite{aihara_2018} on Subaru or the Large Synoptic Survey Telescope\cite{Oguri_2010} (LSST). 
Follow-up confirmation of the lenses and subsequent monitoring can be conducted on smaller aperture facilities.\cite{courbin_2018} 
Forecasts for lensed supernovae suggest that $\sim$500 will be detected by LSST.\cite{goldstein_2017} 
Using current error-budgets, a 1\% precision measurement of H$_0$ requires gravitational time-delay measurements for 40 systems -- which can be a combination of quasars and supernovae. 

\begin{figure} %%%%%%%%%%%%%%%%%%%%%%%%%%%%%%%%%%%%%%%%%%%%%%%%%%%%%%%%%%%%%%%%%%%%%%
\centering
\includegraphics[width=1.0\textwidth]{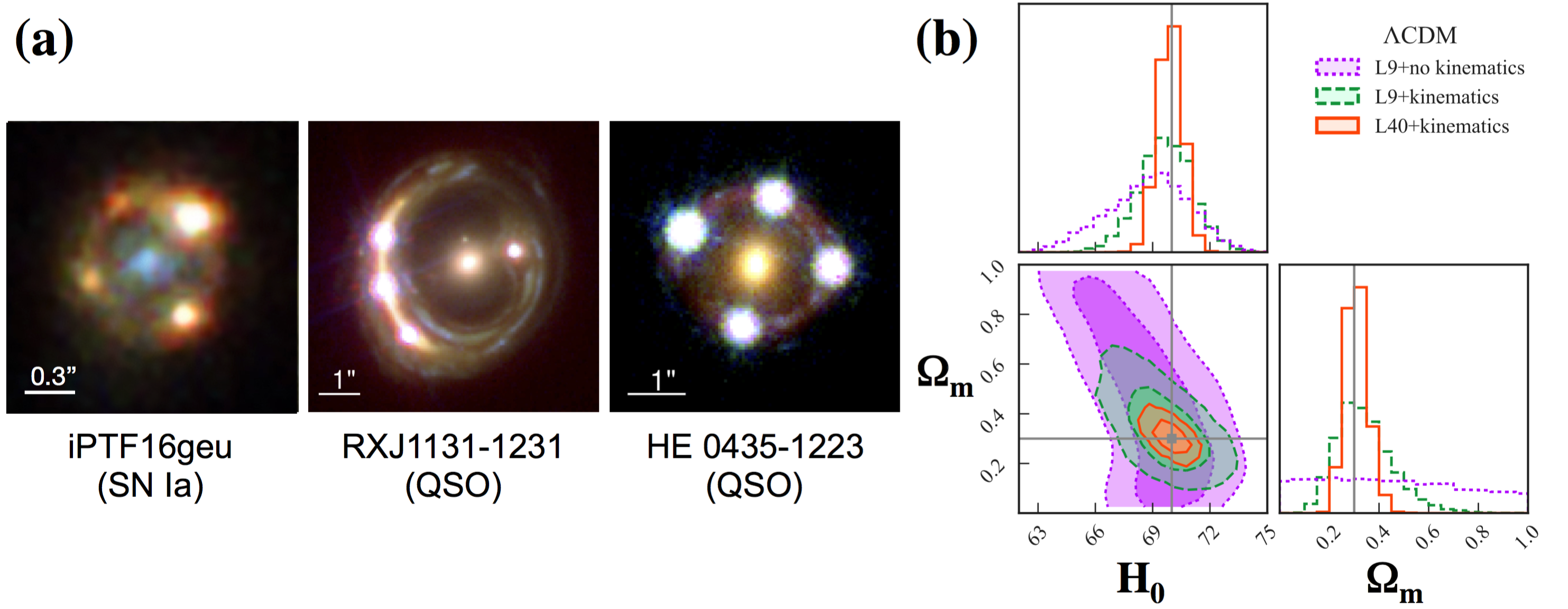} \label{fig:lensing}
\vspace{-11pt}
\caption{(a) HST images of quadruple lenses; lensed Type Ia supernova (SN Ia), iPTF16geu,\cite{goobar_2017} and examples of two lensed quasars (QSO), RXJ1131-1231 and HE0435-1223.\cite{birrer_2019}  
Imaging of this quality or better is required for future lens discoveries.
(c): Comparison of the cosmological constraints in $\Lambda$-CDM from 9 lenses (purple), 9 lenses with high resolution kinematics (green), and 40 lenses with kinematics (orange).\cite{shajib_2018} 
A 1\% H$_0$ provides a tight constraint on $\Omega_{m}$.}
\end{figure} %%%%%%%%%%%%%%%%%%%%%%%%%%%%%%%%%%%%%%%%%%%%%%%%%%%%%%%%%%%%%%%%%%%%%%

%%%%%%%%%%%%%%%%%%%%%%%%%%%%%%%%%%%%%%%%%%%%%%%%%%%%%%%%%%%%%%%%%%%%%%%%%%%%%%%%%%%%%%%%%%%%%%%%%%%%%%%%%%%%
\section{H$_0$ via Standard Sirens} \label{sec:sirens}

Gravitational Wave (GW) signals act as {\it standard sirens} and provide a third independent route to H$_0$.\cite{schutz_1986}
The observable quantities from a GW signal are: the amplitude, $h$, the GW frequency, and the chirp rate and these correspond to three physical parameters of chirp mass, frequency, and the distance.\cite{holz_2005}
To obtain a luminosity distance, one needs either a detection in three GW-detectors or a detection in two GW-detectors and an electromagnetic (EM) counterpart for source localization. 
In either case, the GW-detectors give a measure of $h$ in two polarizations with the third measurement used to disentangle the anistropy due to the antenna pattern (requiring the sky position) from the the specific radiation pattern for the GW event (due to the orbital inclination).
The trigonometric dependence on inclination can be solved for from $h$, but is degenerate for some angles (which increases uncertainty for an individual source).
Together, these parameters determine the distance, where the primary source of uncertainty flows directly from the uncertainty in the wave amplitude, $h$.  
In the coming years, typical instrumental uncertainties from LIGO+VIRGO will be $\sim$5\%, with 1\% accuracy eventually possible.\cite{abbott_2017,chen_2018}

With source localization (either with a third GW detection or electromagnetic counterpart), the only other data required to measure H$_0$ is the redshift of the host galaxy. 
Host galaxies in the Hubble Flow are expected to be faint and the GW sources will likely be distributed across the sky. 
Sufficient sources for a 5\% measure of H$_0$ are anticipated within 5 years of sustained LIGO/VIRGO operation.\cite{chen_2018}
The first kilonova event provided a 10\% estimate of H$_0$, with much of the uncertainty coming from the inclination;\cite{abbott_2017} thus, achieving a 1\% H$_0$ measurement will require redshifts of $\sim$25 GW host galaxies.

\section{Recommendations} 
The combination of a percent-level determination of H$_0$ and the Cosmic Microwave Background (CMB)  provides constraints on the nature of dark energy, the physics of neutrinos, the spatial curvature of the Universe, and has the potential to reveal ``new physics'' with confidence (Fig.~2b). 
While great progress has occurred in the preceding decades, the community still lacks clarity in the the value of H$_0$. 
The coming era of ELTs has the potential to change how H$_0$ is measured, providing three parallel paths that can reach 1\%. 
The critical new capabilities are summarized below. \\
\noindent $\bullet$~ A two-step distance ladder that reaches galaxies in the Hubble Flow via requires deep imaging at high resolution with good image quality. Both JWST and 30-m class facilities are capable of providing this, especially given progress made toward precision photometric work with current MCAO systems.\cite{massari_2016,turri_2016} \\ 
\noindent $\bullet$~ Achieving a accurate distances demands a more sound foundation for the distance scale. Eclipsing binaries have provided the such a foundation in the Large Magellanic Cloud, but spectroscopic follow-up has been limited in Andromeda and other Local Group galaxies at similar distances (e.g., M\,33, IC\,1613, among others). The throughput of 30-m class facilities equipped with moderate to high resolution spectrographs (R$>$25,000) will provide the cruical spectroscopic monitoring for targets discovered from smaller-aperture time-series imaging. \\
\noindent $\bullet$~ Multiple imaged lensing systems of variable sources are anticipated from the deep, time series images produced via the Large Synoptic Survey Telescope (LSST). Long term monitoring of the variable systems can typically be accomplished on modest-aperture telescopes. Open access community broker services, like ANTARES\cite{saha_2014}, are also key to the identification and follow-up of the best sources for these measurements. \\
\noindent $\bullet$~ Gravitational lensing time delay measurements require precise mass modelling. This is obtained via high angular resolution spectro-photometric observations at 0.01$''$-0.02$''$ to resolve the separation between lenses and take a census of the matter field in the vicinity of the lens. These requirements are beyond what is achievable with current instrumentation, including JWST, but will be met by adaptive-optics enabled 30-meter class telescopes. \\
\noindent $\bullet$~ Current forecasts for GW sources anticipate sufficient sources to be detected for a 5\% measure of H$_0$ over five years of LIGO operation.\cite{chen_2018} While the gravitational wave signal, itself, encodes much of the information required for H$_0$ both source localization and host redshifts will be necessary to break the degeneracy in the wave pattern with its observational constraints. Source localization requires relatively modest apertures, but measuring host redshifts will often require the sensitivity of 30-m class facilities.

\pagebreak

%\bibliographystyle{jhep}
%\bibliography{bib.bib}

\providecommand{\href}[2]{#2}\begingroup\raggedright\endgroup

\end{document}